# Aspherical PIC code (APIC) for modeling non-spherical dust in plasmas using shape-conforming coordinates

R.D. Smirnov and S.I. Krasheninnikov

*University of California San Diego, Department of Mechanical and Aerospace Engineering, La Jolla, CA 92093-0411, USA*

**Abstract**

The 2D3V Aspherical Particle-in-Cell (APIC) code is developed for modeling of interactions of non-spherical dust grains with plasmas. It simulates the motion of plasma electrons and ions in a self-consistent electric field of plasma-screened charged dust particle. Due to absorption/recombination of plasma particles impinging on the grain surface, they transfer charge, momentum, angular momentum, as well as kinetic and binding energy, creating currents, forces, torques, and heat fluxes to the grain. The values of such physical parameters determine dust behavior in plasma, including its dynamics and ablation, and can be used in various plasma studies and applications, such as dusty plasmas, fusion devices, laboratory experiments, and astrophysical research. Obtaining these physical values for select non-spherical shapes of conducting dust grains is the main goal of the APIC code simulations.

1. **Methods**

The APIC code uses electrostatic Particle-in-Cell (PIC) [1,2] method for modeling of plasma particles motion in a simulation domain. The PIC method treats a small volume in a phase space of plasma species as a single macro-particle, for which corresponding equations of motions are solved. The macro-particles have the same charge-to-mass ratio, $q/m$, as the corresponding plasma species, while the number of the simulated macro-particles can be set according to available computational resources.

In electrostatic PIC method, the electric force on the plasma particles is calculated by solving the Poisson's equation on a spatial grid and then differentiating the electrostatic potential to find local electric field components at the grid nodes. As the plasma macro-particles positions generally don't coincide with the grid nodes, the spatial distribution of the plasma charge density specified at the grid nodes is computed by particle charge weighing procedure, where each macro-particle's charge is distributed to nearby grid nodes assuming some shape of the particle's volumetric charge distribution. Similarly, components of the electric field are found by spatial interpolation of the field values at the nearby nodes to each macro-particle position.

The 2D3V APIC code can model rotationally symmetric systems that include a conductive non-spherical dust grain in the middle. A constant magnetic field $B$ parallel to the rotational symmetry axis can be present in the modelled system. In this case, Lorentz force acting on plasma particles is also calculated in APIC code. In the electrostatic PIC, it is assumed that plasma currents do not affect the magnetic field significantly, requiring plasma pressure to be much smaller than magnetic field pressure.

The boundaries of the system are selected to correspond to surface of the grain and a closed surface external to the grain. Plasma particles are injected into the system with a given velocity distribution function (in a basis case, unperturbed Maxwellian with a prescribed particle density and temperature) and can leave the system through the outer boundary. Plasma particles intersecting the boundary corresponding to the dust grain surface are assumed to transfer their charge, momentum, and energy to the grain locally at the point of impact. In this way currents, forces, torques, and heat fluxes from plasma to the grain are calculated. Both boundaries are assigned a fixed potential as boundary conditions for solution of the Poisson's equation.

2. **Model implementation**

Currently dust shapes corresponding to equipotential surfaces of combination of a point charge and an electric dipole in vacuum are implemented in APIC. The equipotential surfaces are described as

$$\Phi = \Phi_0\left(\frac{1}{r} + \frac{\alpha \cos\theta}{r^2}\right), \Phi_0 = \frac{Q}{4\pi\varepsilon_0 R_0}, \alpha = \frac{p}{QR_0}, \quad (1)$$

where $Q$, $p$, and $R_0$ are the dust grain's charge, electric dipole moment, and characteristic radius, respectively; $r \geq 0$ and $0 \leq \theta \leq \pi$ are, respectively, the normalized radial distance and azimuthal angle relative to the symmetry axis in polar coordinates with the origin inside the grain. To significantly reduce errors associated with the electric field calculations at the grain surface we introduce dimensionless dust shape conforming curvilinear coordinates $(x_1, x_2)$ in APIC code as follows:

$$x_1 = -\frac{1}{r}\left(1 + \frac{\alpha \cos\theta}{r}\right), x_2 = \cos\theta. \quad (2)$$

Inverse transformation of the coordinates can also be expressed analytically:

$$r = -\frac{1 + \sqrt{1 - 4\alpha x_1 x_2}}{2x_1}, \theta = \text{acos}\, x_2. \quad (3)$$

We note that the conforming coordinates $(x_1, x_2)$ are non-orthogonal. The surface of the dust grain corresponds to constant $x_1 = -1$ and the outer boundary to some small constant value $-1 < x_1 < 0$. Then, the differential element of surface area corresponding to a constant $x_1$ can be expressed as

$$dA = 2\pi R_0^2 r^2 \sqrt{1 + (1 - x_2^2)\left(\frac{\alpha}{r + 2\alpha x_2}\right)^2}\, dx_2, \quad (4)$$

and the differential volume element as

$$d\Omega = \frac{2\pi R_0^3 r^4}{1 + 2\alpha x_2/r} dx_1 dx_2, \quad (5)$$

which are used in APIC code for the charge weighing and the plasma particle injection purposes.

The Poisson's equation for potential $\varphi$ in the domain with plasma charge density $\rho$ is

$$\triangle_{x_1, x_2} \varphi(x_1, x_2) = -\frac{\rho(x_1, x_2)}{\varepsilon_0}, \quad (6)$$

where Laplace operator in the curvilinear coordinates has form

$$\triangle_{x_1,x_2} = \frac{(r+2\alpha x_2)^2 + \alpha^2(1-x_2^2)}{r^6}\frac{\partial^2}{\partial x_1^2} + \frac{1-x_2^2}{r^2}\frac{\partial^2}{\partial x_2^2} - \frac{2\alpha(1-x_2^2)}{r^4}\frac{\partial^2}{\partial x_1 \partial x_2} - \frac{2x_2}{r^2}\frac{\partial}{\partial x_2}. \quad (7)$$

For simplicity of interpretation and possibility to generalize simulations to other curvilinear coordinates, we describe dynamics of plasma particles using dimensional cylindrical coordinates $(R, Z)$, where

$$R = R_0 r\sqrt{1 - x_2^2},\ Z = R_0 r x_2. \tag{8}$$

Corresponding plasma particle equations of motion are

$$\begin{cases} \dfrac{dR}{dt} = V_R,\ \dfrac{dZ}{dt} = V_Z, \\ \dfrac{dV_R}{dt} = \dfrac{q}{m}(E_R - V_\Theta B) + \dfrac{V_\Theta^2}{R},\ \dfrac{dV_Z}{dt} = \dfrac{q}{m}E_Z,\ \dfrac{dV_\Theta}{dt} = \dfrac{q}{m}V_R B - \dfrac{V_R V_\Theta}{R}. \end{cases} \tag{9}$$

The electric field components in the cylindrical coordinates can be found from a solution of Eq. (6) for potential $\varphi(x_1, x_2)$, as follows

$$E_R = -\dfrac{\partial x_1}{\partial R}\dfrac{\partial \varphi}{\partial x_1} - \dfrac{\partial x_2}{\partial R}\dfrac{\partial \varphi}{\partial x_2},\ E_Z = -\dfrac{\partial x_1}{\partial Z}\dfrac{\partial \varphi}{\partial x_1} - \dfrac{\partial x_2}{\partial Z}\dfrac{\partial \varphi}{\partial x_2}. \tag{10}$$

We note that in Eq. (10) we find the derivatives of the potential numerically at the grid nodes with subsequent interpolation to a plasma particle position, while the coordinate derivatives are found analytically for arbitrary spatial coordinates. The electric field calculated on the boundary corresponding to the conducting grain surface is used to find the surface charge density from the normal field component and the electric force acting on the grain. The plasma currents, drag force due to plasma particles absorption, torques, and heat fluxes are computed, correspondingly, as plasma particles charge, momentum, angular momentum, and energy transferred to the dust in unit time interval. The APIC code also has capability to model Boltzmann distributed electrons, in which case electron motion is not simulated and, instead, electron density is calculated as $n_e = n_0 \exp(-e\varphi/T_{e0})$.

The equations of motion (9) are integrated in time using explicit leapfrog first order method. The simulation grid is selected to be uniform in curvilinear coordinates $(x_1, x_2)$, on which finite difference derivatives are calculated to the second order precision. Then, a linear system of the finite difference equations corresponding to Poisson's equation (6) is solved using LU matrix decomposition routines from BLAS/LAPACK libraries. The code is written in Fortran and parallelized using OpenMP techniques.

### 3. Simulations example

To illustrate the capabilities of the code we conducted preliminary simulations of non-spherical dust with $R_0 = 10\ \mu m$, $Q = -1.5 \cdot 10^4\ e$, $p = 0.036\ e \cdot m$ in a plasma with unperturbed density $n_0 = 5.5 \cdot 10^{17}\ m^{-3}$, electron to ion mass ratio $m_e/m_i = 0.1$, and electron and ion temperatures equal $T_{e0} = 1.0\ eV$ and $T_{i0} = 0.1\ eV$, respectively.

In Fig.1, the simulated electric force, the drag force due to plasma particles absorption by the grain, and their sum are plotted in time. One can see that there is a very small electric drag forces on the grain from plasma associated with its shape asymmetry. The existence of such force, in particular, the electric component, was theoretically predicted by considering asymmetric scattering of plasma particles by the electric dipole field of the grain in approximation of very large Debye length $\lambda_D \gg R_0$ [3]. However, the APIC simulations go beyond the theoretical approximation by including the finite plasma screening effects and the drag force dust to plasma particles impinging on the grain. The obtained simulation results also predict the flipping of the electric force sign when the plasma screening of the grain is considered.

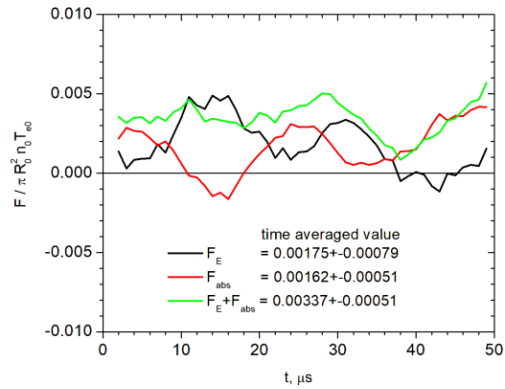

Fig. 1 *Simulated electric force, plasma drag force, and their sum acting on the non-spherical dust in the plasma.*

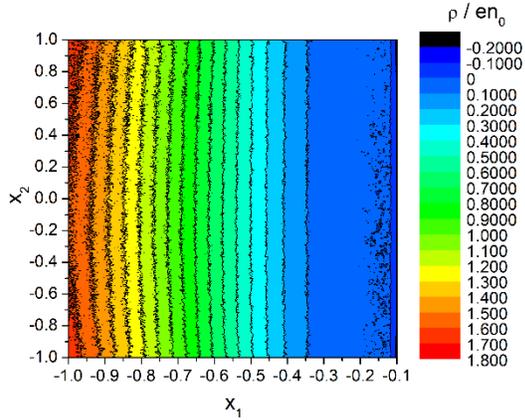 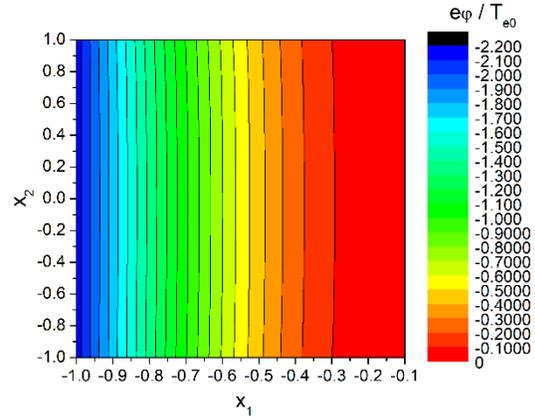

Fig. 2 *Simulated distribution of the plasma charge density.*

Fig. 3 *Simulated distribution of the electric potential.*

The effects of plasma screening is illustrated in Fig. 3, where the distribution of the plasma charge density $\rho$ is plotted. One can see in this figure, that the positive charge accumulates, in particular, near the ends of the grain ($x_2 = \pm 1$) with slightly larger positive charge values near the 'dull' end ($x_2 = +1$). In Fig. 3 the distribution of the electric potential is plotted. The non-sphericity of the plasma screening manifests itself by slightly more positive values of the potential near the 'dull' end ($x_2 = +1$), as compared to the 'sharp' end ($x_2 = -1$) for the same values of $x_1$. We should note that test modeling of the spherical dust shapes in the same plasma with APIC code shows no forces acting on the grain as expected.

### 4. Conclusion

The Aspherical Particle-in-Cell (APIC) code has been developed for modeling the interactions of non-spherical dust grains with plasmas. The code uses dust shape conforming coordinates allowing for precise simulation of very small electric forces acting on the grain. The obtained with APIC simulations preliminary results signify possibility of existence of the electric and the plasma absorption drag forces on non-spherical dust grain, arising due to asymmetries of the dust-plasma interactions, even when the unperturbed plasma is not flowing.

[1] C. K. Birdsall and A. B. Langdon, "Plasma Physics via Computer Simulation", (New York: McGraw-Hill, 1985).
[2] R. W. Hockney and J. W. Eastwood, "Computer Simulation using Particles", (New York: Taylor & Francis Group, 1988).
[3] S. I. Krasheninnikov and R. D. Smirnov, "On the force exerted on a non-spherical dust grain from homogeneous, stationary, non-magnetized plasma", arXiv preprint (2023) arXiv:2305.01846.